# Amino Acids Stabilizing Effect on Protein and Colloidal Dispersions

Ting Mao[1†], Xufeng Xu[1†], Pamina M. Winkler[1†], Cécilia Siri[1], Ekaterina Poliukhina[1], Paulo Jacob Silva[1], Zhi Luo[2], Quy Ong[1*], Alfredo-Alexander Katz[3*], Francesco Stellacci[1,4,5*]

1. Institute of Materials, Ecole Polytechnique Fédérale de Lausanne (EPFL), Lausanne, Switzerland
2. Department of Biomedical Engineering, Southern University of Science and Technology; Shenzhen 518055, Guangdong, P.R. China
3. Department of Materials Science and Engineering, Massachusetts Institute of Technology (MIT), Cambridge (MA), USA
4. Bioengineering Institute, Ecole Polytechnique Fédérale de Lausanne (EPFL), Lausanne, Switzerland
5. Global Health Institute, Ecole Polytechnique Fédérale de Lausanne (EPFL), Lausanne, Switzerland

[†]These authors contributed Equally
*Corresponding authors: quy.ong@epfl.ch, aalexand@MIT.EDU, francesco.stellacci@epfl.ch

**Abstract**

Despite being used for decades as stabilizers, amino acids (AAs) remain mysterious components of many medical and biological formulations. Hypotheses on their role vary ranging from hydrotropic to protein-specific effects (stabilization against misfolding). Here, we deduce that AAs possess a new and broad colloidal property by finding that stabilizing effect of the AAs is comparable on dispersion of various proteins, plasmid DNA, and non-biological nanoparticles. The interactions among colloidal particles in dispersion are carefully evaluated by the second osmotic virial coefficient ($B_{22}$) and the potential of mean force. We propose a theoretical framework that explains the stabilization as the effect of weakly interacting small molecules with patchy nanoscale colloids. We validate it through quantitative comparison with experimental data by comparing equilibrium dissociation constants for AA/proteins obtained either by fitting the $B_{22}$ data with this theory or experimentally. We find excellent quantitative agreement (e.g. proline/lysozyme 1.18 and 2.28 M, respectively) and indeed that the interactions are very weak. The theory presented implies that (i) charged AAs will be effective only for proteins of opposite charge; (ii) short peptides composed of *n* AAs will be as or more effective than *n* separate AAs; (iii) any small molecule weakly interacting with nanoscale colloids that increases the solvation of the surface will have a stabilizing effect. The experimental evidences corroborate all three predictions. Much like the ionic strength of the solution is commonly reported, our results imply that the same should be done for the small molecules, as they also affect fundamentally colloidal properties. As an example, we show that AAs vary the cloud point of a lysozyme solution by as much as 4 K.

**Introduction**

The stabilizing effect of amino acids (AAs) is widely employed in developing protein-based medical formulations[1]. To illustrate how widespread the use of AAs is, we provide here a few examples. Cuvitru, a common immunodeficiency drug, contains glycine to prevent aggregation and denaturation of immunoglobulin G (IgG)[1]. A mixture of glycine and histidine is added to formulations of recombinant proteins (Factor VIII in Kovaltry[2], Factor IX in BeneFIX[3]) used for hemophilia treatment. Proline reduces the formation of dimer and polymer aggregates of human IgG. Hence, proline is used to improve stability as well as clinical tolerability of intravenous immunoglobulin (IVIg) solutions[4]. By stabilizing the monomeric form of the protein, proline markedly prolongs the activity and purity of the IgG product for up to 3 years[5]. Furthermore, in Privigen[6] and Hizentra[7], proline is used as an excipient at a concentration of 250 mM. In biology, the role of AAs in the cytosol of cells has been a topic of investigation for years. It was reported that almost all water-stressed organisms reduce the aggregation of proteins in cells by raising the concentration of AAs such as proline and glutamic acid[8]. *Escherichia Coli* regulates cytosolic proline concentration to inhibit the initial aggregation process of cellular retinoic acid-binding proteins[9].

Despite such widespread use, we are far from understanding the nature of AAs' stabilizing effect on dispersions. In fact, we do not even know if it is of biological nature as opposed to being a generic colloidal property. Many are the proposed mechanisms. A large body of literature shows that AAs can help stabilize proteins against misfolding, thus implicitly postulating that AAs stabilization effect is derived from a specific protein-AA interaction[10–12]. It has been postulated that AAs may have hydrotropic effects, *i.e.* they would stabilize protein solutions above a threshold concentration (minimum hydrotrope concentration) by interacting with the protein surface and possibly masking the exposed hydrophobic parts[13,14]. In addition, AAs (proline, alanine, hydroxyproline, etc.) have been shown to have strong effect on the water H-bonding structure[15–17]. To the best of our knowledge, no study exists on the effect of AAs in stabilizing dispersions of non-biological colloids and most explanations of their role consistently imply some form of biological effect. Furthermore, we are not aware of any predictive theory that would be able, from an independent measurement, to derive whether a specific AA or an additive can stabilize a given protein.

Here, we study the effect of AAs on a wide range of colloidal dispersions (proteins, nanoparticles, and plasmid DNA) by measuring the second osmotic virial coefficient ($B_{22}$) and the potential of mean force. In all cases, we find that $B_{22}$ increases with the addition of AAs indicating more stable dispersions. Changes in $B_{22}$ are detectable at concentrations as low as 10 mM with no concentration threshold and at protein to AAs stoichiometric ratios as low as 1:7. We postulate that the effect is due to weak AA–colloids interactions that in turn modulate the colloid self-interactions. We present a theory that effectively captures this phenomenon. The theory considers proteins or generally colloids as patchy particles, and the number of effective patches is determined by how many of these sites are not blocked by bound AAs. The number of blocked patches is assumed to follow a Langmuir isotherm, whereas the stability of the colloids is evaluated at the mean-field level. The theory is validated through quantitative comparison with experimental data and by using it in a predictive manner to find other biological (peptides) and non-biological molecules that effectively stabilize protein dispersions.

**Results and Discussion**

To investigate whether the ability of AAs to stabilize protein dispersions is a colloidal or a protein-specific property, we explored the effect of proline on gold nanoparticles (AuNPs) coated with a mixture of hydrophobic and hydrophilic ligands. The latter are nanoscale colloids that have significant differences when compared to proteins. For example, no misfolding is possible in AuNPs. Specifically, we used AuNPs with a diameter of 3.5 nm coated with $(56.5 \pm 4.3)\%$ of mercapto-undecane sodium sulfonate and $(43.5 \pm 4.3)\%$ of octanethiol (true stoichiometric ratio, see **Figure S1** for the experimental determination). The nanoparticles in water form a stable suspension, as determined by experiments performed in analytical ultracentrifugation-sedimentation equilibrium (AUC-SE), showing angular velocity dependent equation of state (EOS) curves[18] (**Figure S2**). We used a recently developed cryo-electron microscopy method[19] to derive the particles' potential of mean force (PMF). In **Figure 1a**, we show PMF curves in water at varying particle concentrations. In all three curves, there is a minimum in the potential at ~7 nm due to the formation of aggregates; at ~20 nm we find the corresponding energy barrier caused by electrostatic repulsion. This interpretation of the curves is consistent with recent literature[20]. The minimum in the potential depends on the concentration of the particles in its magnitude becoming deeper as the concentration increases, but its position stays constant. The barrier height becomes smaller and the barrier position moves to shorter distances as the concentration of the particles increases. The energy difference between the potential minimum and the barrier height varies from 3 to 4 $kT$ as the concentration decreases. These values suggest that the aggregation is dynamic in these suspensions. For comparison, in **Figure 1b** we present the PMF for the same nanoparticles in a solution containing proline at a concentration of 2.0 M. As we vary nanoparticle concentrations, the minimum of the potential shows the same behavior as observed in the absence of proline. In stark contrast, however, the concentration dependence of the energy barrier is very different when proline is present. In this case, the barrier height is basically independent of the nanoparticle concentration and its position varies only minimally with it. In **Figure 1c**, **1d** and **1e**, we present the same data shown in **Figure 1a** and **1b** in a different way, namely, we compare directly suspensions at the same nanoparticle concentration with and without proline. It is immediately evident that proline has no clear effect on the minimum of the potential, but it increases the barrier height. As expected, the effect diminishes with decreasing concentration as the suspensions behave closer to their ideal limit. We used the same approach to investigate the effect of proline on ferritin, a protein that is simple to image in cryo-EM. As shown in **Figure 1f**, the PMF for ferritin shows also a minimum and a small aggregation barrier and the addition of 1.0 M proline renders the PMF much closer to that of ferritin at a lower concentration. Overall, this set of data indicates that proline stabilizes colloidal suspensions as its addition renders the PMFs closer to that of suspensions at lower concentrations. More importantly, we find that the proline effect appears to be very similar in the case of a protein and a nanoparticle, suggesting a colloidal effect and not a protein-specific one.

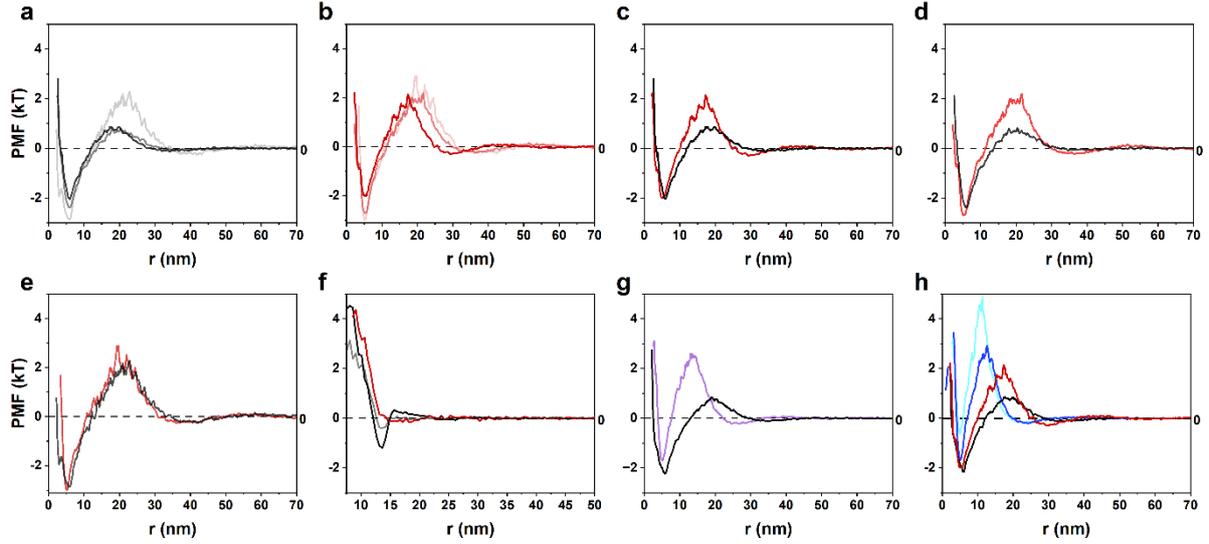

**Figure 1**. PMFs of Au nanoparticles and ferritin in different solvent additives and concentrations. a) PMF of AuNPs dispersion in water for different particle concentrations (in units of $particles/m^3$): $6.39 \times 10^{22}$ (black), $5.14 \times 10^{22}$ (dark grey), and $3.61 \times 10^{22}$ (light grey) (b) PMF of the same AuNPs dispersions in 2M proline for different particle concentrations (in units of $particles/m^3$): $7.01 \times 10^{22}$ (solid red), $4.89 \times 10^{22}$ (medium red), and $3.51 \times 10^{22}$ (light red); (c-e) the proline effect (red) in comparison to pure water (black) of the AuNPs of the similar number concentration; (f) PMF of ferritin in 50 mM phosphate buffer in number density of $13.3 \times 10^{22}$ (black) and, $3.3 \times 10^{22}$ $particle/m^3$ (dark grey), as well as with the addition of 1.0 M proline in number density of $10 \times 10^{22}$ $particle/m^3$ (solid red); (g) PMF of AuNPs in water in number density of $6.35 \times 10^{22}$ $particle/m^3$ (solid black) in contrast with the addition of 2.0 M TME in number density of $7.59 \times 10^{22}$ $particle/m^3$ (purple); (h) PMF of Au nanoparticles in water (solid black), 2M proline (red), 0.67 M triproline (blue) and 0.5M tetraproline (cyan) in number density of $6.32 \times 10^{22}, 7.01 \times 10^{22}, 8.95 \times 10^{22}$ and $6.67 \times 10^{22}$ $particle/m^3$ respectively. In case of the PMFs shown in (g) and (h) we should note that the barrier position moves to shorter distances and its height increases consistent with an effect of the molecules added that stabilize the monomeric form of the nanoparticles against its aggregated one.

One common way to evaluate the stability of a compound in a dispersion is through the equation of state (EOS)[18,21], shown below in Eq. (1):

$$\frac{\Pi}{kT} = \rho_2 + B_{22}\rho_2^2 + B_{23}\rho_2\rho_3 + \cdots \quad (1)$$

where $\Pi$ is the osmotic pressure, $\rho$ is the number density and $kT$ is the product of the Boltzmann constant $(k)$ by the temperature $(T)$. In this equation, $B$ indicates the virial coefficient with the number subscripts indicating the component of the dispersion, 1 the solvent, and 2 and 3 as the main and minor solute. Therefore, $B_{22}$ is the osmotic second virial coefficient that measures the self-interaction among the main solutes. When all virial coefficients have a value of 0, the EOS becomes equivalent to that for the ideal gas. A virial expansion is needed to represent realistic dispersions. We used analytical ultra-centrifugation (AUC)[22,23] and self-interaction chromatography (SIC)[24,25] to measure $B_{22}$. The choice of two independent methods based on different physical principles was made to avoid any possible measurement-related bias or error[26]. We measured the change of $B_{22}$ ($\Delta B_{22}$) as a function of the concentration of AAs in dispersion. A positive change ($\Delta B_{22} > 0$) indicates that in the dispersion the net interactions become more repulsive (i.e., that the difference between repulsion and attraction grows in favor of the former). As expected, we find that adding salt or a suitable polymer to a protein dispersion leads to a $\Delta B_{22} < 0$ due to increased electrostatic screening or depletion forces respectively (**Figure S3**). As shown in **Figure 2**, the addition of proline to a dispersion of lysozyme, bovine serum albumin (BSA), apoferritin, or fused in sarcoma (FUS) low complexity domain (LCD) (residues 1–267 [27]) (**Figure 2a, 2c** and **2d** respectively) leads to a positive change in $B_{22}$ ($\Delta B_{22} > 0$), despite an obvious increase in the ionic strength due to the zwitterionic nature of proline at buffer pH 7.0 (isoelectric point of proline: 6.3). Similar results were observed when proline was added to a dispersion of plasmid DNA (**Figure 2e**), or to the same of nanoparticles described above (**Figure 2f**). Proline was used as a representative amino acid, yet all AAs that we tested yielded $\Delta B_{22} > 0$ for lysozyme (**Figure 2b**). It should be stressed that there is a significant difference between proteins and nanoparticles. The former can change conformation (through misfolding or smaller transient conformational changes), but the latter cannot[28–30]. The fact that $\Delta B_{22} > 0$ upon AAs addition for both cases suggests that conformational changes are not the reason for this phenomenon. In a recent publication[18], we have shown that lysozyme and BSA dispersions are solutions while apoferritin is a suspension, hence the thermodynamic status of the dispersion seems not to be relevant for this effect to occur. Apoferritin (molecular weight $M_w$=480,000 Da) and plasmid DNA ($M_w$=2,600,000 Da) are much larger than lysozyme ($M_w$=14,300 Da) and BSA ($M_w$=66,000 Da), indicating an effect that happens over a wide span of sizes. FUS-LCD is an intrinsically disordered protein, while the other proteins we tested are folded ones. Yet, they all change $B_{22}$ in a similar manner upon proline addition.

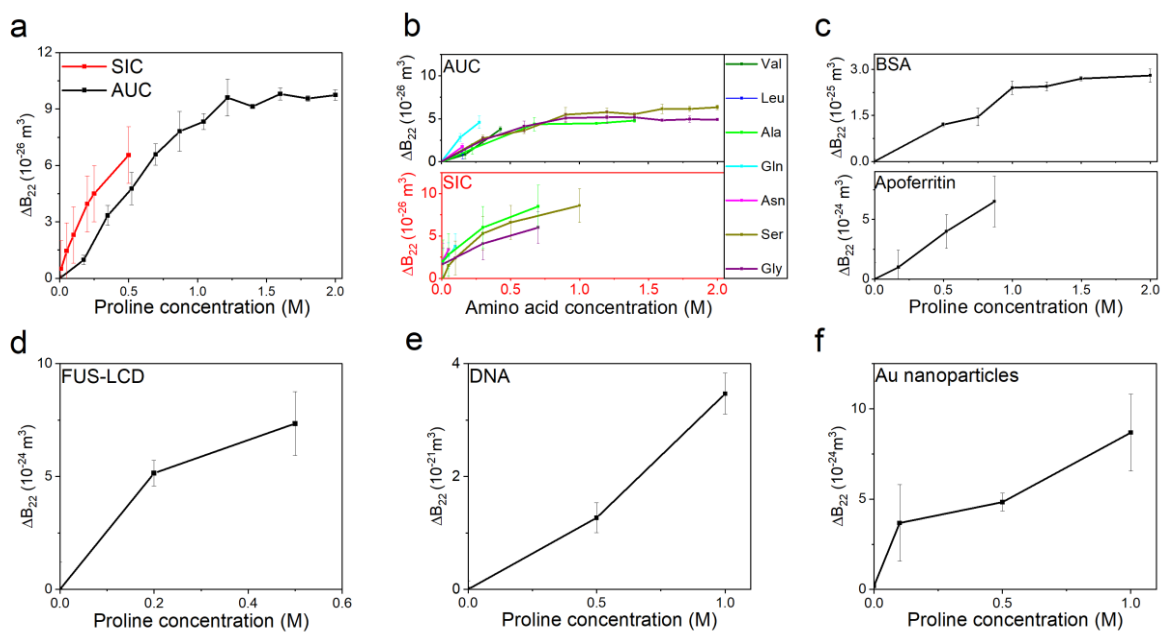

**Figure 2.** Change of the second virial coefficient $B_{22}$ ($\Delta B_{22}$) for different self-interactions as a function of added amino acids (AAs). By applying Sedimentation diffusion Equilibrium Analytical Ultra-Centrifugation (SE-AUC, in black) and Self-Interaction Chromotography (SIC, in red) we quantifiy the influence of increasing proline concentration (a) and different amino acids (b) on the self-interaction of lysoyzme. By applying SE-AUC we measure the modulation upon the addition of proline on the self-interacting Bovine Serum Albumbin (BSA) and apoferritin (c), FUS-LCD (d), DNA (e) and Au NPs (f). Note that SIC has been established with lysozyme being manually grafted to the column and thus only the self-interaction of lysozyme can be probed. To prevent the clogging of the SIC column the effect of small molecules is typically investigated in the concentration range below 1.0 M, especially for less soluble molecules.

As directly demonstrated by SIC (**Figure 2b**) we measure positive changes in $\Delta B_{22}$ at AAs concentrations as low as 10 mM corresponding to a molar ratio of lysozyme to AA of about 1:7 (shown for glutamine in **Figure S4**, representative for all AAs). We performed a classical hydrotrope assay[31], in which the solubility of a hydrophobic compound (fluorescein diacetate FDA) in water is measured as a function of hydrotrope concentration. As shown in **Figure S5**, we did not find a threshold concentration for proline as the minimum hydrotrope concentration. Furthermore, we tested $\Delta B_{22}$ dependence on the absolute concentration of the AA or on the stoichiometric ratio between the AA and the protein. We found that $\Delta B_{22}$ depends only on the absolute AA concentration and not on that of the proteins (**Figure S6**).

We propose a theoretical framework to explain all the experimental observations here presented. It is based on the hypothesis that weak interactions between the AAs and the colloids in dispersion lead to a time-averaged screening of colloid–colloid attractive interactions. For the simplicity of presentation, we discuss below the case of AAs interacting with proteins, but the model is general and is developed for small molecules interacting with nanoscale colloids in a dispersion. Consider a set of $n$ colloidal particles with $z$ attractive patches where they can interact with other identical colloidal particles. Given the coordination number $z$, we can think of this problem as that of colloids in a lattice with coordination given by $z$. We set the volume of a lattice point to be approximately the volume of the colloids $a^3$ for simplicity. The coordination number depends on the number of patches and is an open variable, but it will be in general a number of order 10 or less (e.g., the coordination number for a face center cubic lattice would be 12, but it is 2 for divalent patchy particles). The coordination number sets the fictitious lattice on which we will work, but it is not important for the rest of our calculations nor has further implications.

The free energy (per site) can be constructed directly from a mean field (Bragg-Williams) approach to yield

$$\frac{\Delta G}{kT} = \phi \ln \phi + (1-\phi)\ln(1-\phi) + \chi \phi (1-\phi) \qquad (2)$$

where $\phi$ is the volume fraction of the colloids within the solvent. The interaction is completely captured by the incompatibility parameter $\chi$ (normally referred to in soft matter as the Flory $\chi$ parameter)

$$\chi = \frac{z}{kT}\left(\epsilon_{CS} - \frac{\epsilon_{SS}+\epsilon_{CC}}{2}\right) \qquad (3)$$

where the different $\epsilon's$ represent an effective interaction energy (per patch) between either colloid–colloid (cc), solvent–solvent (ss), or the cross-interaction colloid–solvent (cs). As can be seen from Eq. (3), increasing (or decreasing) the incompatibility parameter $\chi$, which dictates the enthalpy of mixing, can be accomplished via increasing (decreasing) the strength of the interactions $\epsilon's$, or by changing the effective coordination of the colloids $z$. We believe that the mechanism by which AAs modulate colloid-colloid interaction is a competition-type of interaction, meaning AAs decrease $z$ by "blocking" attractive patches and hydrating them. The second virial coefficient can be extracted from the expansion of the free energy, and is written as

$$B_{22} = a^3(1 - 2\chi) \quad (4)$$

Note that in the absence of attractive interactions (i.e., $\chi \to 0$), the excluded volume becomes the lattice site volume. More generally, we should evaluate the second virial as $B_{22} = -\int (e^{U/kT} - 1) d\vec{r}$, where $U$ is the intermolecular potential. It is important to note that within the model we are studying the attractive interaction parametrized by $\chi$ is assume to be a contact interaction of very short range, *i. e.* a *contact interactions* occurring only upon contact of the particles, and this gives rise to the interacting part scaling with $a^3$. This is a good approximation in the limit that the interactions are short ranged. However, if long-range electrostatic repulsive interactions are present, one would need to add a constant to the virial potential and express it as:

$$B_{22} = a^3 \left(1 + \frac{1}{a^3}\Delta B_{22}^{elec} - 2\chi\right) \quad (5)$$

where $\Delta B_{22}^{elec} = -\int_a^\infty (e^{U_{elec}/kT} - 1) d\vec{r}$. For a pure Coulomb potential, one finds that $\Delta B_{22}^{elec}$ does not converge, but it does for a Debye-Yukawa potential, which is expected under the conditions typically encountered in colloidal solutions. As can be seen from (5), this generalized form of the second virial coefficient is similar to the original case where no electrostatic contributions were assumed. One only needs to rescale the effective size of the colloid by the full repulsive excluded volume and rescale the attractive part accordingly. Thus, for simplicity we will assume from here on expression (4) without losing generality, with the understanding that $a^3$ corresponds to an effective excluded volume and not just the steric contribution. As will be seen below, this rescaling does not affect any of our results.

Now we turn our attention to the effect of AAs or similar small molecules. We assume that AAs adsorb onto some patches of the proteins, this is in agreement with literature reports that have observed the adsorption (i.e. preferential hydration) of some AAs onto protein surfaces[32–35]. The adsorption is transient but it equilibrates rapidly since the barriers to adsorb and desorb are small, and the diffusion constant of the AAs is large compared to that of the proteins themselves [36,37]. Thus, we can think of this system as consisting of an adsorption onto a surface with $Nz$ adsorption sites, where $N$ is the number of proteins, and each has $z$ patches to adsorb onto. We will assume for simplicity that one patch can only adsorb one AA or small molecule. This is not a strict condition and can be easily relaxed by assuming each patch has a given number of adsorption sites. We can model this adsorption process using the Langmuir adsorption isotherm, where we can compute the fractional coverage (fraction of adsorbed sites) as

$$\theta = \frac{Kc}{1+Kc} \qquad (6)$$

where $c$ the concentration of AAs and $K$ is the equilibrium constant of binding between AAs and attractive patches. These quantities can be measured or computed.

The effective coordination is then given by $z = z_0(1-\theta)$, with $z_0$ being the original number of patches in the absence of AAs. Finally, the second virial coefficient becomes

$$B_{22} = a^3\left(1 - 2\frac{z_0}{kT}(1-\theta)F(\{\epsilon\})\right) \qquad (7)$$

where $F(\{\epsilon\})$ is a function that only depends on the original interaction parameters and is independent of AAs concentration.

The second virial coefficient in the absence of amino acids (when $\theta = 0$) can be written as

$$B_{22}{}^o = a^3\left(1 - 2\frac{z_0}{kT}F(\{\epsilon\})\right) \qquad (8)$$

and the change in the virial coefficient upon addition of AAs is then simply given by

$$\Delta B_{22} \sim 2a^3 \frac{z_0}{kT}\theta F(\{\epsilon\}) \qquad (9)$$

Inserting the expression of $B_{22}{}^o$ in terms of $F(\{\epsilon\})$ in $\Delta B_{22}$ from (8) and explicitly considering the fraction of sites we find that the change of the second virial coefficient is an expression independent of the interaction energies

$$\Delta B_{22} = (a^3 - B_{22}{}^o)\frac{Kc}{1+Kc} \qquad (10)$$

This expression is very insightful since it shows that the scale of change in the second virial is dictated by the difference between the maximum excluded volume the colloid can have given by $a^3$ (stemming from purely repulsive interactions, steric + electrostatic, as discussed above), and the measured/actual excluded volume in the absence of AAs, i.e. $B_{22}{}^o$ that does include the effect from attractive patches. The change is regulated by the number of blocked attractive sites/patches given by the Langmuir expression where the equilibrium constant $K$ can be extracted by fitting. It is important to mention that we are treating all patches equally. This condition can be relaxed as shown in the supplementary material and does not affect the results we are presenting. Furthermore, this relationship allows for the quantification of the strength of the interaction that will lead to noticeable changes, these will happen when the product of *K* and *c* is in the range of 1.

Using Eq. (10) we could fit all the data we presented on the change of the experimental second virial coefficient, $\Delta B_{22}$, as a function of concentration $c$ of AAs. For proline, glycine, serine, and alanine we had enough data points to achieve reliable fits. We started by fitting the proline data as they had the most data points. We did so by floating the $K$ and $a$ parameters and obtained an $a^3$ of $1.7 \times 10^{-25} \pm 2.2 \times 10^{-26} \, m^3$ and $K$ of $0.85 \pm 0.26 \, M^{-1}$ (**Table 1**, detailed fitting in **Figure S7**). We first comment on $a$ whose value implies a diameter of 5.5 nm that is in good agreement with the hydrodynamic diameter of lysozyme that we measured with dynamic light scattering (DLS) to be 4.1 nm (**Figure S8**) also matching the reported literature value[38]. We then constrained $a$ to this value and fitted the data for the other AAs. We obtained the values for $K$ that are shown in **Table 1**. We also measured the interaction of AAs with lysozyme experimentally by using intrinsic fluorescence experiments[39] (see **Table 1**). For proline–lysozyme we measure an equilibrium dissociation constant $K_D$ of $2.28 \pm 0.07 \, M$ in great agreement with the one deduced by fitting our $\Delta B_{22}$ data, $K_D = 1/K = 1.18 \pm 0.36 \, M$. We observe that the experimental data and the fitted data never differ by more than a factor of 2 (see **Table 1**), a remarkable agreement given the uncertainties involved. Intrinsic fluorescence experiments are sensitive to changes in the tryptophan fluorescence of the protein, which can be induced by direct binding of AA or by a conformational rearrangement triggered by the AA interaction[39,40]. There are three tryptophan residues on a lysozyme surface. This experiment does not exclude binding in far-away surface sites; but given the weak nature of the interaction we can exclude that many more AAs will bind to lysozyme from simple thermodynamic considerations. We also performed the same analysis for BSA, and $a^3$ of $1.8 \times 10^{-24} \pm 1.0 \times 10^{-25} \, m^3$ was found. This value corresponds to a diameter of ~12 nm that is in good agreement with ~ 9.8 nm obtained by DLS (**Figure S8**). The $K_D$ that leads to the best fit of $\Delta B_{22}$ data is $2.00 \pm 0.52 \, M$ (**Figure S9**). We measured proline–BSA interactions by a competitive binding assay that gave a $K_D$ value of $3.0 \pm 1.0 \, M$, which is again in excellent agreement.

**Table 1**. Comparison of the equilibrium binding constant $K_D$ obtained.

| Amino acids | Proline | Glycine | Serine | Alanine |
|---|---|---|---|---|
| *From fit of $\Delta B_{22}$ Plot** | | | | |
| $K$ (M$^{-1}$) | 0.85±0.26 | 0.33±0.05 | 0.42±0.03 | 0.39±0.04 |
| $K_D$ (M) | 1.18±0.36 | 3.03±0.46 | 2.38±0.17 | 2.56±0.26 |
| *From Intrinsic Fluorescence Measurements*** | | | | |
| $n$ | 0.96 ± 0.03 | 0.76 ± 0.05 | 0.85 ± 0.08 | 0.92 ± 0.03 |
| $K_D$ (M) | 2.28 ± 0.07 | 5.93 ± 0.12 | 4.98 ± 0.36 | 1.76 ± 0.04 |

* Equilibrium constant of binding ($K$) and the calculated dissociation constant ($K_D = 1/K$) by fitting the SE-AUC data with Eq. (10) ($a$ = 5.5 nm for the fitting)

**$K_D$ measured by intrinsic fluorescence experiments for lysozyme with 4 different AAs, being proline, glycine, serine and alanine. The number of binding sites $n$ has been determined using the Eq. (11) (for more details see **Methods and Materials**) obtained from the fitting the intrinsic fluorescence measurements.

There are many direct consequences of this theory. The first is that AAs with charged side chains will have a stabilizing effect only on proteins of the opposite charge and this is confirmed by our finding as shown in **Figure 3a** and **3b**. Second, in a first approximation assuming the interaction scales linearly with the number of AA per oligomer, the theory implies that short peptides composed of *n* AAs should produce a $\Delta B_{22}$ on a colloidal solution that is equivalent to the $\Delta B_{22}$ produced if *n* AAs were added separately to the solution. As shown in **Figure 3c** when we measure $\Delta B_{22}$ for peptides formed by three or four prolines, respectively, we find that the obtained values of $\Delta B_{22}$ are basically the same ones as generated by proline itself (see also SIC results, **Figure S10a**). We can further confirm this by studying the PMF for gold nanoparticles when these peptides of proline are added. As shown in **Figure 1h** tetra-proline appears to be even more effective than proline in stabilizing particles solutions. Third, the theory does not imply any special property for AAs. It only requires molecules to have weak interactions with the colloids. Thus, we selected a common molecule:

1,1,1-Tris(hydroxymethyl)ethane (TME) (see insert in Fig. 3d), that is highly water-soluble (solubility limit > 2.5 M in water[41]) but has an exposed methyl group. As shown in **Figure 3d** TME produces a $\Delta B_{22}$ for lysozyme similar to that one produced by AAs (see also SIC results, **Figure S10b**). It is also worth noting that in this TME case (**Figure 1g**) the changes in PMF for AuNPs are significant.

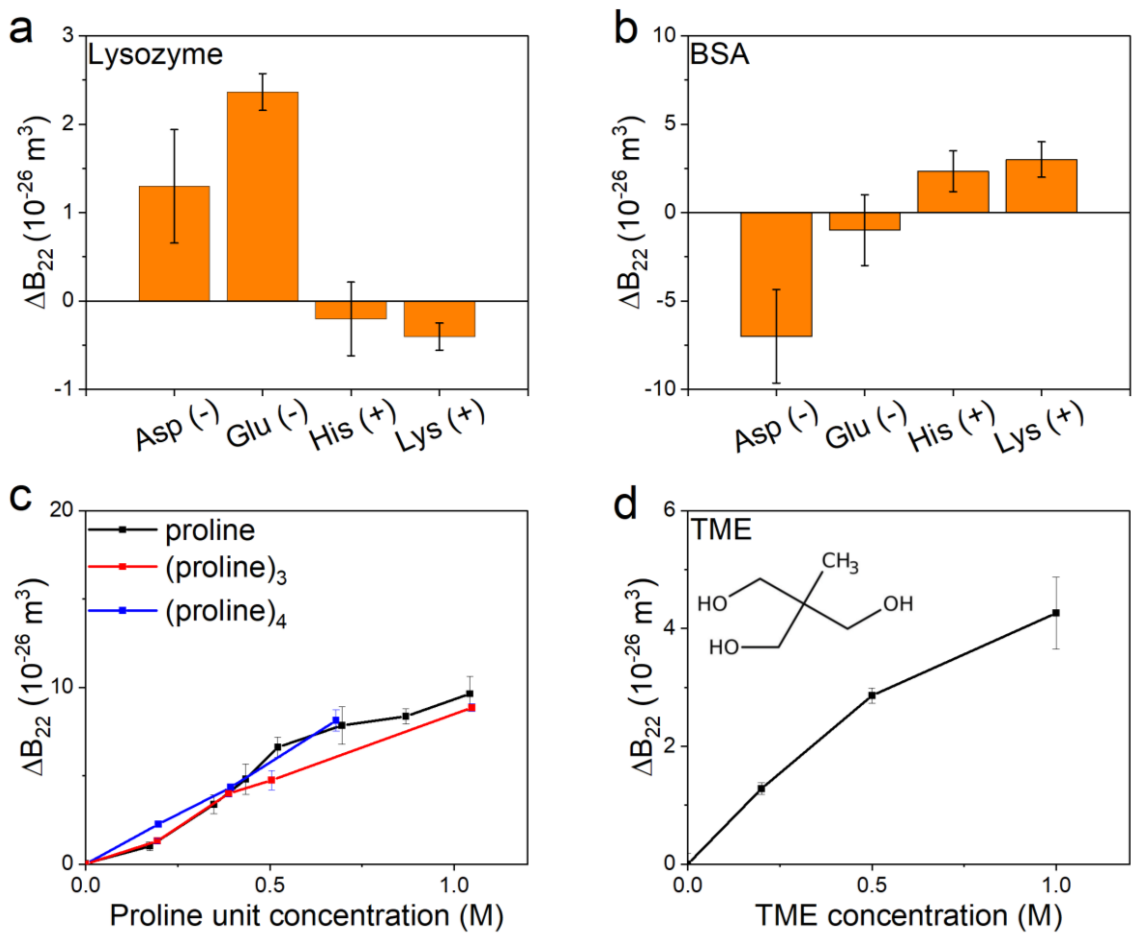

**Figure 3**. Change of the second virial coefficient $B_{22}$ ($\Delta B_{22}$) by charged amino acids, poly-amino acids and the small molecule TME. By applying SE-AUC we quantify the effect of two positively (histidine and lysine) and two negatively (aspartic acids and glutamic acid) charged amino acids on the positively charged lysozyme (a) and on the negatively charged BSA (b) self-interaction at neutral pH. By applying SE-AUC we quantifiy the self-interaction of lysozyme in presence of poly-proline peptidues (tri-proline (proline)₃ and tetra-proline (proline)₄) in compairson to proline (c) and of TME (d).

Protein−protein interactions are key to the determination of the coexistence curve on protein phase diagrams, where the equilibrium between protein-rich and protein-diluted phases is determined. A simple, commonly studied protein phase diagram is that of lysozyme[42–44]. Thus, if AAs modulate protein-protein interactions, they should have an effect on the protein phase diagram by moving the phase-equilibrium lines upward (destabilizing effect of AAs) or downward (stabilizing effect of AAs). We performed cloud point measurements for lysozyme and obtained the phase diagram shown in **Figure 4a**. The addition of 250 mM proline moves the equilibrium line in the phase diagram down as much as ~4 K for protein concentrations lower than 7.0 mM. The critical temperatures ($T_{cr}$) and critical lysozyme concentrations ($C_{cr}$) were found by fitting the coexistence curves by the following equation (11):

$$Cloud\ point\ =\ T_{cr}\left(1-\alpha\left(\frac{C_{cr}-C_p}{C_{cr}}\right)\right)^{\frac{1}{\beta}} \qquad (11)$$

where $C_p$ is the protein concentration, $\beta = 0.33$ is the critical exponent, $\alpha$, $T_{cr}$, and $C_{cr}$ are the adjustable parameters[43]. Using the cloud point equation above to fit our experimental data in **Figure 4a**, we found that the critical temperature $T_{cr}$ and the critical concentration $C_{cr}$ to be 22.9 °C and 15.3 mM in 50 mM phosphate buffer, while they become 17.1 °C and 14.6 mM, respectively, upon the addition of 250 mM proline.

The strength of stabilization depends on the proline concentration. This is shown in **Figure 4b** where the $\Delta T$ for the phase line at lysozyme concentration of 7.0 mM is plotted against proline and many other AAs concentrations. This plot resembles the $\Delta B_{22}$ plot shown in **Figure 2b**, although with some noticeable deviation. This is to be expected as cloud point measurements are performed on a two-phase equilibrium while $\Delta B_{22}$ measures the thermodynamic state of a single phase. Yet, the cloud point measurements confirm that AAs lead to measurable changes in $\Delta T$ at concentrations as low as 4.2 mM and at a protein to AA stoichiometric ratio as low as 1:0.5 (**Figure S11**). Furthermore, we find that the $\Delta T$ change on the phase line scales rather well with the concentration of the peptide units when proline is replaced by tri- or tetra-proline (**Figure S12**), confirming the results obtained from $\Delta B_{22}$. Recently, protein phase separation has been intensely studied in relation with the formation of membrane-less organelles in cells. In a recent paper, we show that AAs have strong effects in relation to the formation of these organelles, always suppressing phase separation[45]. Another direct consequence of increased stability in solution is the cross-interaction between proteins, their ability to bind to each other and their propensity to form aggregates. Recently, we have demonstrated[46] that AAs modulate protein cross-interactions at millimolar concentrations.

For example, 10 mM of glutamine reduces the binding constant of lysozyme to a whey protein isolate by one order of magnitude, in other cases we have found changes of as much as five orders of magnitude.

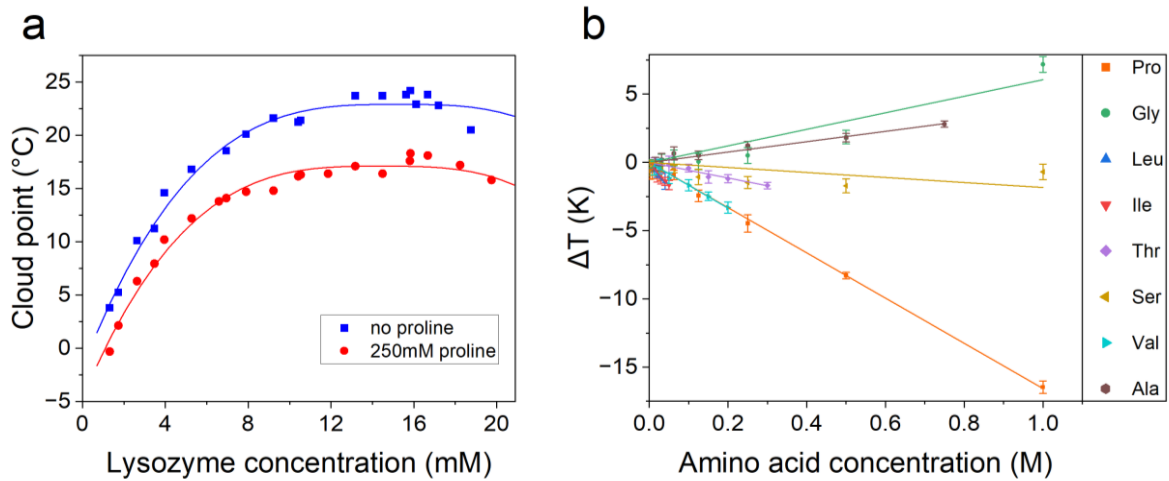

**Figure 4:** Cloud point measurements of lysozyme. (a) parallel comparison of cloud point temperature of lysozyme in various concentration in 50 mM phosphate buffer and with the addition of 250 mM proline; (b) changes of cloud point temperature of lysozyme in 50 mM phosphate buffer with addition of different amino acids in varied concentrations, together with a simple linear extrapolation for each amino acid to help read the scatters.

The experimental data and theory presented suggests that proline has a generic colloidal effect. The $B_{22}$ for a lysozyme solution in 50 mM phosphate buffer (pH 7) was measured with AUC to be $+1.2 \times 10^{-26} \ m^3$. The addition of 0.3 M of NaCl brings it to $-10.5 \times 10^{-26} \ m^3$, as expected salt lowers the $B_{22}$ leading to net attractive interactions ($B_{22} < 0$). The further addition of 1.0 M proline to this solution brings back the $B_{22}$ to the value of $-1.5 \times 10^{-26} \ m^3$ showing that proline can indeed counter the effect of salt. In the case of the addition of 0.2 M NaCl the $B_{22}$ becomes $-7.0 \times 10^{-26} \ m^3$, now the addition of 1M proline brings $B_{22}$ not only back to being positive but also to a value larger that the starting one ($+3.0 \times 10^{-26} \ m^3$), indicating that the effect of proline exceeds that of NaCl at this concentration. It is accepted that salt have a generic colloidal effect that happens through the screening of the electrostatic interactions. We believe that proline has an equally generic colloidal effect that screen attractive interactions and thus can counter the effect of salt (all the data with error bars can be found in **Figure S13** and **Table S1**).

**In conclusion**, in this paper we have shown that weak interactions between small molecules and patchy nanoscale colloids provide sizeable effects on the stability and the self-interactions between these colloids. This becomes especially interesting the case of amino acids interacting with proteins where molar interactions lead to strong changes in the second osmotic virial coefficient and in the potential of mean force. In turn, these changes have large effects on protein phase-diagrams, liquid-liquid phase separation, and binding constants. The effects shown exist for amino acids as well as short peptides, both are abundant in cell cytosols and we believe that their role is underappreciated. It is known that cells under osmotic pressure produce more amino acids, but they also produce more peptides and amino acids when they degrade intracellular proteins. Our data shows that the increase in such intracellular concentration of AAs and peptides will influence all protein-protein interactions.

**Acknowledgments**

Funding: This work was supported by Swiss National Science Foundation, European Union's Horizon 2020 Research and Innovation program under Grant Agreement No. 101017821 (LIGHT-CAP), and Nestle research foundation. Z.L. thanks Dr. Ye Yang, Prof. Stefan Guldin, Dr. Anna Murello, Prof. Paolo Arosio, Dr. Lenka Faltová, Prof. Juan Aragones, and Prof. Francesco Sciortino for helpful discussions. A. A.-K. would like to thank the Michael and Sonja Koerner for their generous support. The Cryo-EM data collection and initial image processing were performed at the Dubochet Center for Imaging Lausanne (a joint initiative from EPFL, UNIGE, UNIL, UNIBE) with the assistance of A. Myasnikov, B. Beckert, S. Nazarov, Inayathulla Mohammed, and E. Uchikawa.

Author contributions: QO, FS conceived and supervised the project. AAK developed the theory. QO, TM, XX, PW, CS, EP designed and carried out the experiments, and analyzed the results. PJS synthesized ligands. All authors discussed the results and co-wrote the manuscript.

Competing interests: The authors declare no other competing interests.

Data and materials availability: All data are provided in the main text or the supplementary materials.

**Materials and Methods**

Hen egg-white lysozyme (14.3 kDa, ≥ 95%) was purchased from Roche. All the AAs were purchased from Thermo Scientific in powder form. FUS-LCD was a gift from Prof. Dufresne[27]. BSA was purchased from Thermo Scientific. Ferritin, apoferritin were ordered from Sigma Aldrich. Poly-proline peptides (tri- and tetra-proline) were purchased from Bachem Americas. 1,1,1-Tris(hydroxymethyl)ethane (TME, 97%) was purchased from ABCR Swiss AG. Plasmid DNA (4207 bp ds-DNA, pIVEX1.3-CAT) was purchased from Biotechrabbit GmbH. The salts for the sodium phosphate buffer were also purchased from Sigma Aldrich in powder form.

**Synthesis of nanoparticles**

789.3 mg of $HAuCl_4·3H_2O$, 64 mL of oleylamine and 80 mL of n-octane were mixed in the 500 mL three-necks round bottom flask. The mixture was left stirring till the solid completely dissolved. The flask was connected to argon flow for 10 minutes followed by a quick injection of 16 ml of tert-butylamine-borane complex (351.3 mg) dissolved in oleylamine to induce the reduction reaction. The reaction was kept under stirring for 1 hour and quenched with 240 ml of ethanol. Nanoparticles were precipitated with centrifugation. Followed by sonication adding fresh dichloromethane then ethanol several times and reprecipitate each time before adding fresh ethanol to remove organic residuals from reaction materials. 90 mg of oleylamine functionalized nanoparticles was dissolved in 30 ml dichloromethane (dissolved with 34.86 mg of sodium mercapto-undecane sulphonate (MUS) and 31.23 ul of octane thiol). The ligand exchange was kept sealed and under stirring for 21 hours. The nanoparticles after ligand exchange were sonicated and precipitated several times with fresh dichloromethane and acetone to remove organic residuals. Then pre-purified nanoparticles were dissolved in ultrapure water using Amicon® 30 kDa MWCO filters for further purification to remove MUS. NMR of nanoparticles is presented in the supporting information (**Figure S1**) to show the quality of purification as well as the final composition of the two ligands functionalized on the nanoparticles.

**Purification of ferritin based on sucrose density gradient ultracentrifugation**

Ferritin, from equine spleen Type I (Sigma Aldrich, in saline solution), known to have a wide distribution of ferritin due to different iron loads, oligomerization or apoferritin impurities[47], was purified using reparative ultracentrifugation. 5-30 wt.% sucrose gradients in Mili-Q water were prepared in six ultracentrifuge tubes (Ultra-Clear, 25x89 mm, SW28, Beckman Coulter, USA) by a Piston Gradient Fractionator (BioComp Instruments, Canada). 400 μL of ferritin solution was loaded onto the gradient followed by centrifuging (Optima XPN-80, Beckman Coulter, USA) in a SW32 rotor followed by centrifugation at 30,000 rpm at 20 °C for 2 h. Then, three fractions were collected in a top-down manner (liquid surface being 0mm), at distances of 20–28 mm (fraction 1), 28–36 mm (fraction 2), and 36–46 mm (fraction 3). Each fraction was purified 5 times with PBS1x in Amicon tubes (100 kDa cut-off, Ultra-15), at 4000 rpm, 4 °C for 30 minutes. Quality of purification were characterized by analytical ultracentrifugation in sedimentation velocity (AUC-SV), see **Figure S14**. For AUC-SV, 380 μL of PBS1x (reference) and 376 μL of the sample were loaded into 12mm cells in an An60-Ti rotor, equilibrated at 20 °C in vacuum for 2 hrs, followed by velocity scans at 15,000 rpm with absorbance profile collected at 280 nm. Data analysis were done in SEDFIT software[48] with maximum entropy regularization at a confidence level of 0.68, final s-resolution of 500 in s-interval from 0-250 S. For the Cryo-EM measurement, fraction 2 was chosen due to its relatively monomeric distribution at around 60 S that is typical for ferritin[47] (**Figure S14**). Concentration of ferritin was measured by Implen NanoPhotometer (Implen, Germany) at 420 nm using an extinction coefficient[49] of $10\ (mg/ml)^{-1}\ cm^{-1}$.

**Cryogenic Transmission Microscopy**

3.5 ul of dispersion was casted onto a previously glow discharged quantifoil grid (Quantifoil® R 1.2/1.3, 200 Mesh, Cu). The grid with solution was blotted with Whatman filter papers on both sides in a vitrobot (Vitrobot Mark IV) under 100% of humidity at 22 °C, followed by immediate vitrification in liquid ethane. Imaging was done at the Dubochet Center for Imagine (Lausanne) in a Titan Krios G4 operating at 300 KV. Tilt series were recorded from -60 to 60 with 2 increments at magnification of 33000 X (camera pixel size of 0.37 nm), defocus of -7 um with a total dose of 120 e⁻ Å⁻². For data processing, pre-aligned .mrc files were compiled from Tomography version 5.16.0 (thermo fisher scientific). Camera Falcon 4i equipped with Selectris X energy filter (slit width 20 eV).

**Column grafting for SIC experiments**

For this work we adapted and optimized the experimental procedure for the custom-made column grafting with lysozyme for our SIC experiments the protocol established by Le Brun et al.[25] For the grafting of the SIC column, a Tricorn 5/50 column (Cytiva, Column Volume of 1.178 ml) was manually grafted with lysozyme (Lys) using as a resin TOYOPEARL®AF Formyl-650M chromatography particles, sodium cyanoborohydride, potassium phosphate and ethanolamine. The standard buffer used throughout the experiments was 50 mM sodium phosphate buffer at pH ≈ 6.9 consisting of monobasic and dibasic sodium phosphate in MilliQ water.

All experiments were performed on SIC columns with a grafted surface coverage of lysozyme of ≈ 45%. The column was packed under pressure with the following flow rates: 0.75 ml/min for 15 minutes, then at 3 ml/min at 15 min and again at 0.75 ml/min for 30 min. The column was stored at 4°C overnight and between experiment days.

**SIC experiments**

To prepare a chromatography column for a CIC experiment, Lys was manually grafted on the column, as outlined in the previous section on **Column grafting for SIC experiments,** and for the optimized signal-to-noise ratio of the elution profile we determined that the protein concentration should be in the range of ~20 mg/ml which was injected in all SIC experiments shown in this work.

SIC experiments were conducted to probe the Lys self-interaction in different solution environments (buffer alone or in the presence of small molecules e.g., amino acids, poly-proline and TME at different concentrations dissolved in buffer) from which the respective values of $B_{22}$ were calculated. Before each measurement series, a column performance test was run with 20% Acetone in MilliQ water. For each run of the experiments, 50 μL of lysozyme at 20 mg/ml was injected. Samples were injected after 10× Column Volume and with a constant flow rate of 0.75ml/min at room temperature. The amino acids tested were proline, glycine, serine, threonine, asparagine and glutamine and their concentration was varied between 5 mM and 1.2 M (considering their respective solubility limit). The upper range limit was set by the fact that towards the solubility limit of an added molecule (e.g., amino acid, poly-proline peptides, TME) the buffer solution becomes turbid which very likely clogs and thus breaks the column. Therefore, to protect the grafted column we set the upper limit to maximally 1.2 M, a range well below the solubility limit of the studied amino acids and small molecules.

**Determination of $B_{22}$ by SIC**

In SIC experiments, one evaluates the interactions between the injected protein in the mobile phase and the immobilized protein grafted on the column in terms of a measured retention volume. To experimentally determine $B_{22}$ one first computes the retention factor $k' = (V_0 - V_R)/V_0$, where $V_0$ is the retention volume of non-interacting species which is calculated before each experiment with the column performance test using 20% acetone in MilliQ water and $V_R$ is the volume required to elute the injected protein in the mobile phase through the grafted column. Then, $B_{22}$ (mol ml g$^{-2}$) can be computed as:

$$B_{22} = B_{HS} - k's\Phi$$

where $B_{HS} = 2\pi r^3/3N_A M^2$ is the excluded volume or hard sphere contribution of the two interacting proteins assuming a spherical shape, $s$ being the immobilization density, i.e., the number of covalently immobilized protein molecules per unit area of the bare chromatography particles and $\Phi = A_S V_0$ is the phase ratio (i.e., the total surface available to the mobile phase protein. To calculate $B_{HS}$, we used the hydrodynamic radius $r = (1.89 \pm 0.03)$ nm and the averaged molecular weight $M = 14300$ g/mol of the self-interacting lysozyme[38], $N_A$ refers to the Avogadro's number. To exchange the units for $B_{22}$ (mol ml g$^{-2}$) which is commonly used in SIC experiments to the one of AUC experiments $B_{22}$ (m$^3$) = $B_{22}(mol\ ml\ g^{-2})\ M^2/N_A 10^{-6}$.

The assumption here is that we measure only two-body interactions, i.e., one injected free protein interacts with only one immobilized protein molecule at a time. This constraint can be guaranteed by controlling the immobilized proteins grafted onto an effectively flat surface being the column. Another assumption is that the injected free Lys interacts only with one immobilized Lys grafted onto the column and not with each other. This can be verified by determining the variation for the calculated $B_{22}$ value measured at a concentration of 5-30 mg/ml of the lysozyme. For the Lys-Lys self-interaction the obtained $B_{22}$ value should remain constant. We determined this variation to be $\sim 3.4 \times 10^{-26} m^3$ for Lys-Lys and considered this variation in our error analysis.

**Sedimentation diffusion Equilibrium Analytical Ultra-Centrifugation (SE-AUC)**

In a typical SE-AUC experiment, a protein solution in phosphate buffer (pH 7, 50 mM) at a typical concentration of 10 mg/ml was mixed with AAs. The final solutions were added into the AUC cells (3 mm pathlength). The sedimentation-diffusion equilibria with a depleted meniscus at a proper angular velocity, 20 °C were reached typically after 24 hrs. The protein concentration gradient was obtained and converted into the EOS curve by the previously established methods[50,51].

**Cloud point measurement**

A 20 mg/ml lysozyme solution in PBS1x was filtered through a 0.45 µm syringe filter and concentrated in Amicon tubes (10 kDa cut-off, Ultra-15) by centrifuging at 5,000 rpm at 4 °C or 20 °C for 50 minutes. Lysozyme concentration was measured by Implen NanoPhotometer (Implen, Germany) at 280 nm using an extinction coefficient of 2.72 $(mg/ml)^{-1} cm^{-1}$. The concentration of the lysozyme stock solution was adjusted to 200 mg/ml. The stock solutions of the amino acids were prepared by dissolving the amino acid powder in water (Milli-Q) until a final volume of 1 ml was reached and adding 1 ml of 2 M NaCl solution in PBS2x. The two-fold serial dilution of amino acids' stock solutions was done in 1 M NaCl solution in PBS1x. Mixing of protein and amino acid solutions as well as transferring of the samples (100 µl) to capillaries (0.3 ml Crimp Neck Micro-Vial, 31.5 x 5.5 mm, clear glass, round bottom) were performed in ThermoMixer F1.5 Eppendorf at 42 °C. The blocks in a CrystalBreeder (Technobis Crystallization Systems, Netherlands) were kept at 42 °C for 10 minutes before the capillaries were placed into the instrument. The transmissivity was calibrated automatically after 5 minutes of temperature equilibration of the sample at 42 °C, followed by cooling down at a rate of 0.2 °C/min under constant nitrogen flow to prevent protein oxidation. The cloud point temperature was registered at 70% transmissivity loss.

**Intrinsic fluorescence measurements of lysozyme**

All fluorescence measurements were performed with the CaryEclipse fluorometer (Agilent, USA) using a 1.0cm quartz cuvette. The fluorescence spectra of Lysozyme (10 µM) with variable concentrations of proline (0-1 M), glycine (0-1 M), serine (0-1 M) and alanine (0-1 M) were measured at a constant temperature (25 °C). The volumes of the samples were adjusted to 1.5 ml with 50 mM sodium phosphate buffer, at pH = 6.9. The samples were excited at 290 nm and the fluorescence was recorded from 310-420 nm with an em/ex slit of 5 nm and a scanning speed of 120 nm/min.

If it is assumed that there are similar and independent binding sites in the protein, the binding constant ($K_D$) and the number of binding sites ($n$) can be determined using the following equation[52,53]:

$$log\left(\frac{F_0-F}{F}\right) = logK_D + nlog(Q) \qquad (11)$$

Where $F_0$ and $F$ are the fluorescence intensities in absence and in presence of amino acids respectively. $Q$ is the quencher concentration, e.g., amino acid. $k_D$ and $n$ can be determined by the intercept and the slope of the linear regression of $log\left(\frac{F_0-F}{F}\right)$ versus $log(Q)$.

**Dynamic Light Scattering (DLS)**

Lysozyme and BSA solutions, about 0.8 mg/ml, in phosphate buffer (pH 7, 50 mM) were filtered (0.22 µm) before DLS measurements using Nano ZS from Malvern. All the DLS measurements were performed with 3 replicates.